\input{aipcheck}

\documentclass[
    ,final            
  ]
  {aipproc}

\layoutstyle{6x9}

\begin{document}

\title{\emph{Fermi} Gamma-ray Space Telescope Observations of Gamma-ray Pulsars}


\author{P.~M.~Saz Parkinson for the $Fermi$-LAT Collaboration\footnote{\tt{http://www-glast.stanford.edu/cgi-bin/people}}}{
  address={Santa Cruz Institute for Particle Physics, University of California, Santa Cruz, CA 95064}
  ,altaddress={e-mail: pablo@scipp.ucsc.edu}
}

\begin{abstract}
The Large Area Telescope on the recently launched $Fermi$ Gamma-ray Space Telescope 
(formerly GLAST), with its large field of view and effective area, combined with its
excellent timing capabilities, is poised to revolutionize the field of gamma-ray astrophysics.
The large improvement in sensitivity over EGRET is expected to result in 
the discovery of many new gamma-ray pulsars, which in turn should lead to fundamental 
advances in our understanding of pulsar physics and the role of neutron stars in the Galaxy.
Almost immediately after launch, $Fermi$ clearly detected all previously
known gamma-ray pulsars and is producing high precision results on these. 
An extensive radio and X-ray timing campaign of known (primarily radio) pulsars 
is being carried out in order to facilitate the discovery of new gamma-ray pulsars. In addition, a 
highly efficient time-differencing technique is being used to 
conduct blind searches for radio-quiet pulsars, which has already resulted in new discoveries.
I present some recent results from searches for pulsars carried out on $Fermi$ data, both blind 
searches, and using contemporaneous timing of known radio pulsars.

\end{abstract}

\maketitle


\section{Introduction}

\begin{figure}
  \includegraphics[height=.4\textheight]{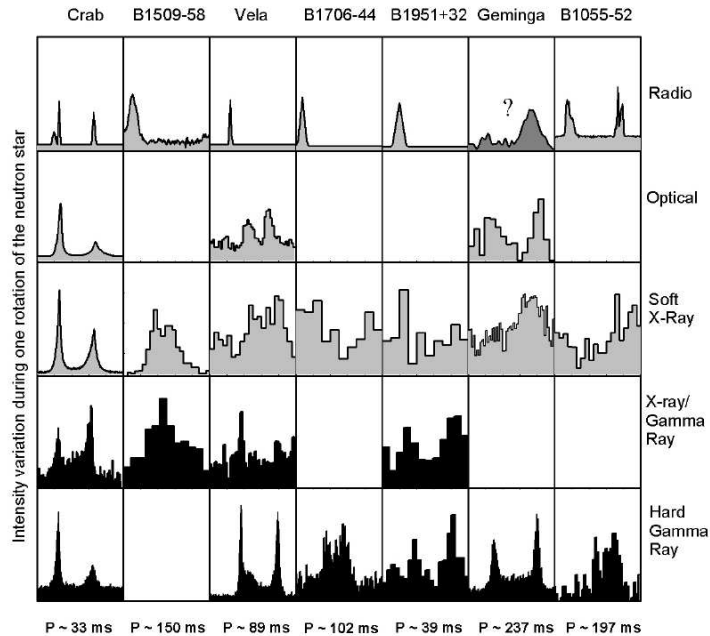}
  \caption{Phase-aligned folded light curves of the CGRO pulsars, at various wavelengths (from \cite{2008RPPh...71k6901T}).}
\label{fig1}
\end{figure}

Among the numerous significant results of the EGRET mission, on the \emph{Compton Gamma Ray Observatory} (CGRO), was the firm detection of 6 gamma-ray pulsars (and marginal detection of at least three additional ones) at energies above 100 MeV. In addition to these, the COMPTEL detector, also on CGRO, detected 
pulsations from PSR B1509--58 up to 10 MeV. More than half the $\sim$300 gamma-ray sources detected by EGRET 
remained unidentified at the end of the mission, and many of these (especially those lying in the Galactic Plane) are 
thought to be pulsars. The light curves, in various energy bands, of these seven CGRO gamma-ray pulsars are shown in Figure~\ref{fig1}. For 
a detailed summary of the EGRET results, see~\cite{2008RPPh...71k6901T}. Of these seven gamma-ray pulsars, six display clear radio pulsations, while 
the seventh, Geminga, is what is known as a "radio-quiet" (or at least "radio-faint") pulsar. Indeed, the entire class of radio-quiet pulsars 
(which until the $Fermi$ era contained this one object) is sometimes referred to as "Geminga-type" pulsars. Differences between the radio and gamma-ray 
beam geometries (e.g. \cite{2008arXiv0812.3931W}) are thought to explain why certain pulsars are radio-loud or radio-quiet. Outer magnetosphere 
models predict a fan-like gamma-ray beam. The much narrower radio beam, on the other hand, can often miss the Earth.

\begin{figure}
  \includegraphics[height=.4\textheight]{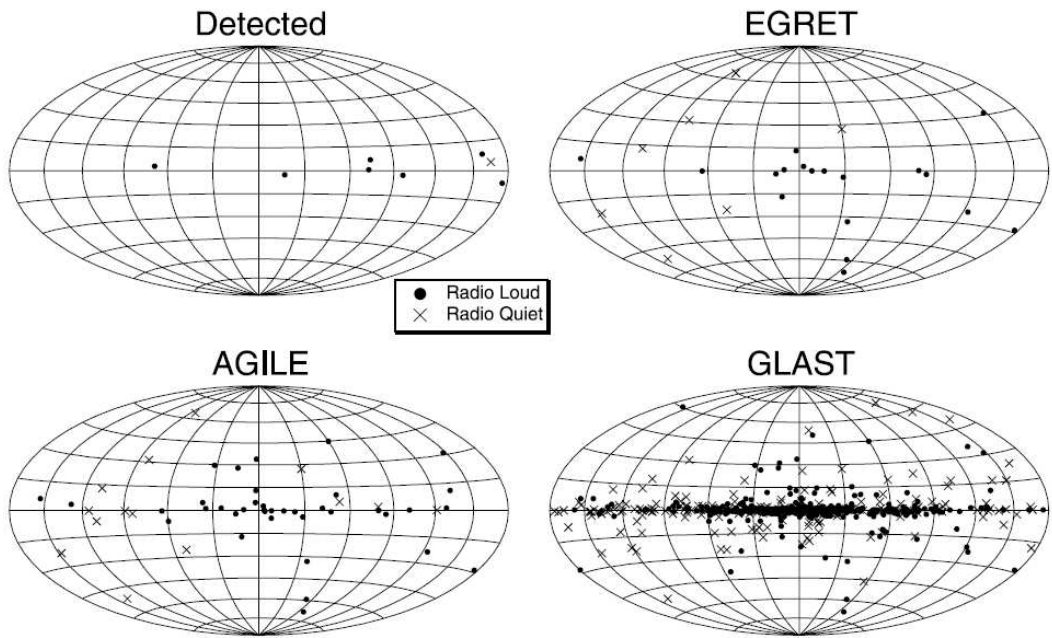}
  \caption{Plots showing the radio-loud (dots) and radio-quiet (crosses) pulsars detected by EGRET (top left), along with the simulated EGRET detections 
(top right), AGILE detections and GLAST ($Fermi$) detections, for a particular population synthesis study (from \cite{2004ApJ...604..775G}).}
\label{fig2}
\end{figure}

Population synthesis models of radio and gamma-ray pulsars (e.g. \cite{2004ApJ...604..775G}) have been used to try to predict the number of radio-loud 
and radio-quiet pulsars that will be detected by the new generation of gamma-ray telescopes (AGILE and $Fermi$). The experimental determination of 
this "Geminga fraction" and a comparison with model predictions will help not only in discriminating 
between different models of gamma-ray emission, but will hopefully also serve to pin down some of the physical (and/or geometrical) parameters of the given 
models. Figure~\ref{fig2} shows, among other things, the large number of radio-loud and radio-quiet pulsars that are expected to be detected 
by $Fermi$, according to one particular population synthesis study~\cite{2004ApJ...604..775G}.

\section{The \emph{Fermi} Large Area Telescope (LAT)}

\begin{figure}
  \includegraphics[height=.3\textheight]{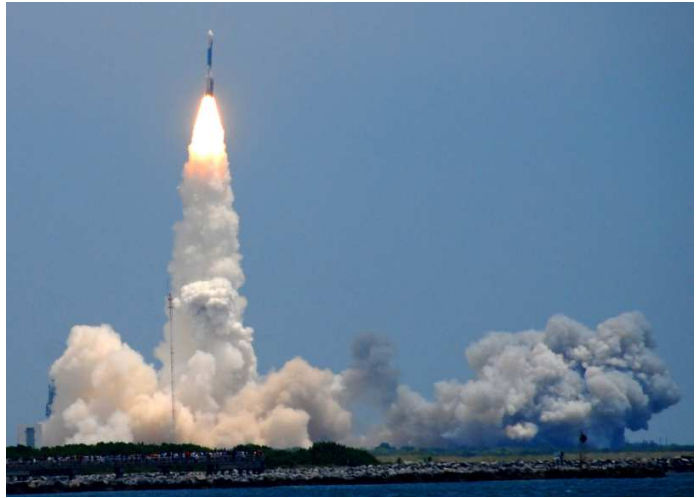}
  \caption{GLAST launch from Cape Canaveral aboard a Delta II rocket. (Credit: M. Ziegler)}
\label{fig3}
\end{figure}

The GLAST (now \emph{Fermi}) satellite, consisting of the Gamma-ray Burst Monitor (GBM) and the Large Area Telescope (LAT), was launched on 11 June 
2008 (see Figure~\ref{fig3}) into a low Earth circular orbit at an altitude of 550 km and an inclination of 28.5$^\circ$. The LAT~\citep{Fermi} is a pair-production telescope with large effective area ($\sim$8000 cm$^2$) and field of view (2.4 sr), 
sensitive to gamma rays between 20 MeV and $>$ 300 GeV. Although its commisioning phase (30 June to 30 July 2008) 
was primarily intended for instrument checkout and calibration, several scientific results have been obtained with these early data. The LAT began normal 
science operations on 11 August 2008, and since then has been observing mostly in survey mode, scanning the entire gamma-ray sky every three hours.
The overall sensitivity of the LAT is $\sim$25 times that of EGRET, while the angular resolution is also significantly improved (it ranges 
from $\sim$3--6$^{\circ}$ at 100 MeV to $\sim$0.1--0.2$^{\circ}$ at 10 GeV). The mission was designed with a five-year lifetime and a goal of at least ten years 
of operations. The scientific goals of the mission include understanding particle acceleration in Active Galactic Nuclei (AGN), pulsars, 
and supernova remnants (SNRs), exploring the high energy emission of Gamma-ray Bursts (GRB), and probing the nature of dark matter (see contribution 
by V. Vitale in these proceedings). 

\section{\emph{Fermi}-LAT searches for Gamma-ray pulsars}

\subsection{Early observations of the EGRET pulsars}

Within the 60-day launch and early operations (L\&EO) period, the six EGRET pulsars were detected. Figure~\ref{fig3} shows 
an early LAT light curve of Geminga. The scale is not zero-suppressed, illustrating the low level of background achieved by the LAT.

\begin{figure}
  \includegraphics[height=.4\textheight]{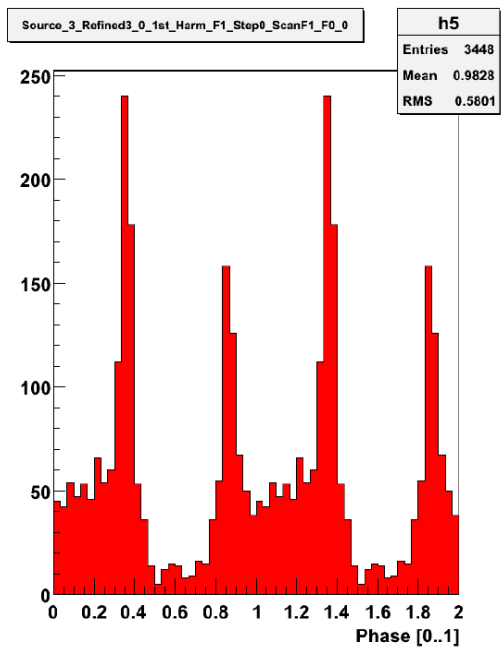}
  \caption{Early (commissioning phase) \emph{Fermi}-LAT light curve of Geminga.}
\label{fig4}
\end{figure}

Early observations of the Vela pulsar, the brightest persistent GeV source in the sky, were used to verify the excellent timing and angular 
resolution of the LAT and to highlight its potential for discovery in the field of pulsar astrophysics (see the contribution by M. Razzano in these 
proceedings, as well as the recently accepted journal article~\cite{2008arXiv0812.2960A}).

\subsubsection{Detection of a glitch in PSR B1706--44}

\begin{figure}
  \includegraphics[height=.3\textheight]{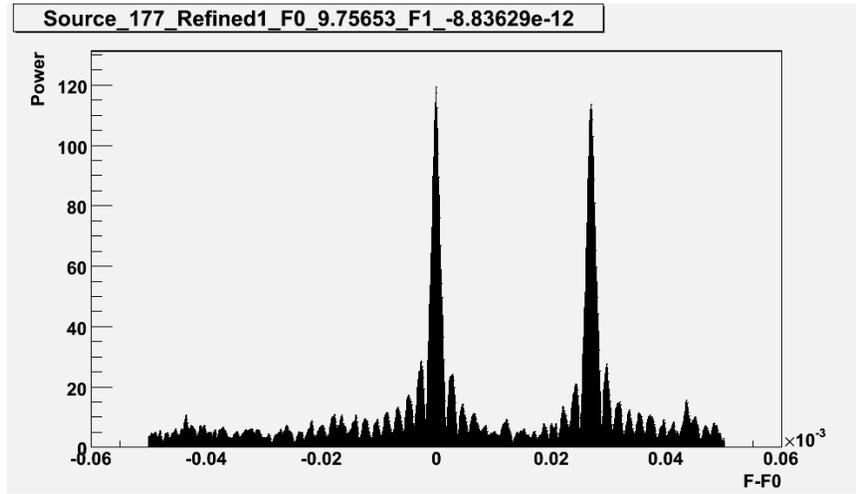}
  \caption{Fourier power spectrum of PSR B1706--44 ($\sim$10 weeks of data) centered around the known frequency of the pulsar, showing 
peaks at both the pre-glitch and post glitch rotation frequencies.}
\label{fig5}
\end{figure}

One of the early surprises in our study of EGRET pulsars came in the form of an apparent glitch in PSR B1706--44. Glitches are discrete changes in the 
rotation speed of a pulsar. They are thought to occur when a sudden transfer of angular momentum takes place from the faster rotating superfluid interior
to the solid crust of the neutron star~\cite{1976ApJ...203..213R}. They are fairly rare, only having been detected in fewer than 50 pulsars, and they 
tend to occur more frequently in young pulsars. The typical fractional increase in rotational frequency is $10^{-10} < \Delta f/f < 5\times10^{-6}$~\cite{2003MNRAS.340.1087K}. A Fourier transform of the data on PSR B1706--44 revealed two highly significant peaks. Figure~\ref{fig4} shows the power spectrum 
centered around the previously known frequency of PSR B1706--44 ($\sim$9.756532 Hz). The second peak is at a frequency 2.8$\times$10$^{-5}$ Hz higher, 
yielding quite a large value of $\Delta f/f \sim 2.9\times10^{-6}$. The data were split in two, to ascertain when the glitch took place and determine 
the best ephemeris for each time period (pre- and post-glitch). The different phase plots are shown in Figure~\ref{fig5}. Due to the low flux of the source 
and the mode of observation of the LAT, the time of the glitch could only be narrowed down to a $\sim$22-hour window: between 08:40 on the 14th of August 
(MJD 54692.36) and 06:00 on the 15th of August (MJD 54693.25). While this pulsar has been known to glitch in the past (e.g. 1995~\cite{1995A&A...293..795J}), 
this represents the first time that a glitch has been detected in gamma rays. Follow-up radio observations are planned, to confirm the glitch.

\begin{figure}[htbp]
\includegraphics[height=.3\textheight]{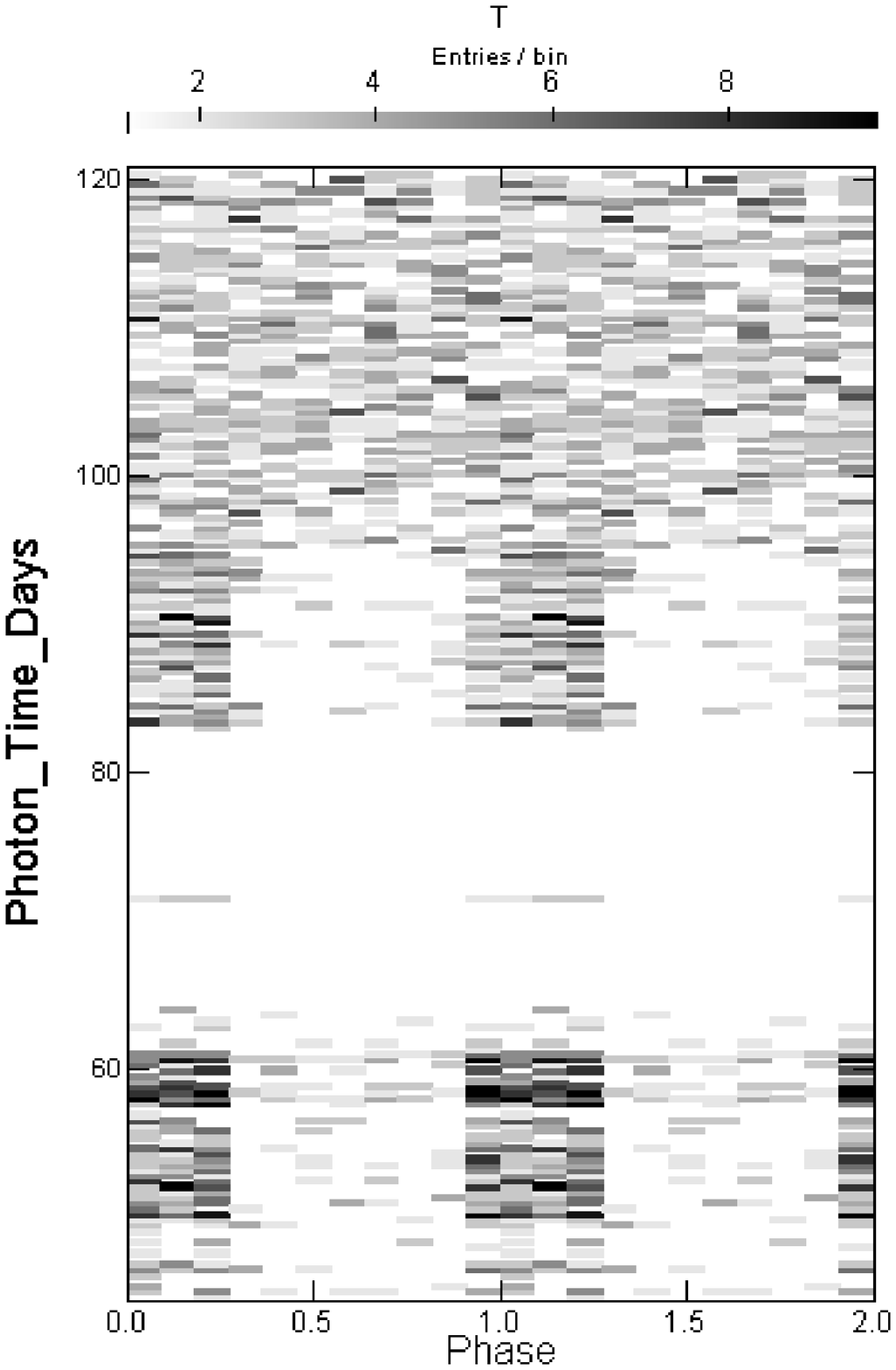}
\includegraphics[height=.3\textheight]{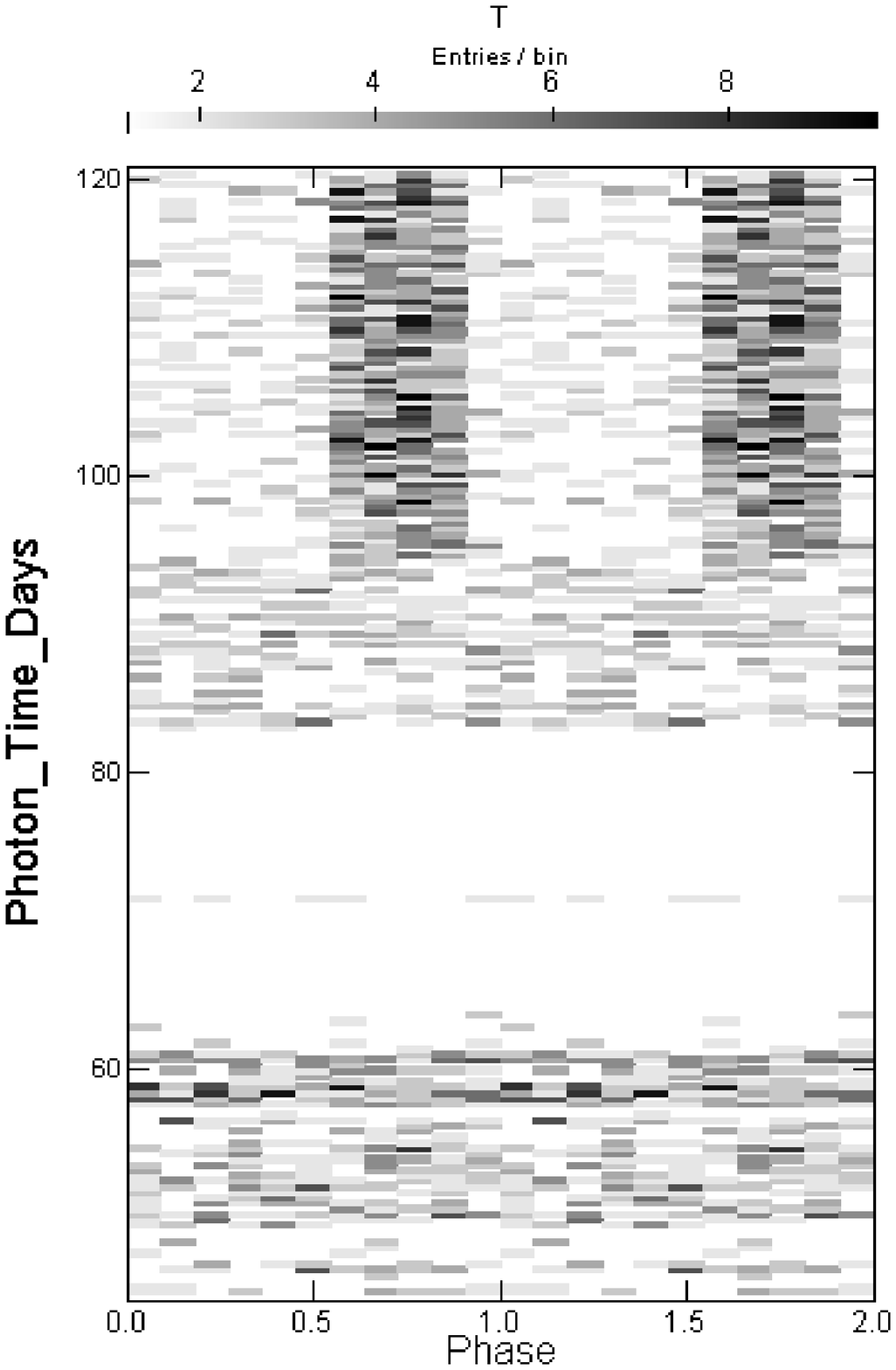}
\caption{PSR B1706--44 phase plots
{\bf Left} -- Using the "pre-glitch" timing solution.
{\bf Right} -- Using the "post-glitch" timing solution.}  
\label{fig6}
\end{figure}

\subsection{Search for gamma-ray pulsations from known pulsars}

Gamma-ray sources are inherently dim and result in very sparse data sets. The detection of gamma-ray pulsations therefore requires 
observations spanning long periods of time (up to years). During such long time spans, many pulsars will experience 
irregularities in their timing behavior, such as "timing noise" or glitches. This makes it necessary to observe pulsars regularly to obtain 
timing solutions valid for the entire period of observation. In anticipation of this, a comprehensive pulsar monitoring 
campaign (also known as the "pulsar timing consortium") was organized prior to the launch of \emph{Fermi} to obtain contemporaneous timing solutions for a 
large (and increasing) number (currently several hundred) of the brightest (as measured by spin-down power) known radio and X-ray pulsars~\cite{smith08}. 
The first results from these efforts came in the form of the detection, 
during the first weeks of the commissioning phase, of pulsations from PSR J2021+3651 (also known as the "Dragonfly", a relatively recently discovered 
radio pulsar coincident with an unidentified EGRET source~\cite{2002ApJ...577L..19R}). The discovery of gamma-ray pulsations from this source was 
recently reported using AGILE observations carried out between 2 November 2007 and 30 June 2008~\cite{2008ApJ...688L..33H}. A detailed publication 
containing the $Fermi$ results is in preparation for submission to the Astrophysical Journal.

\subsection{Blind searches for gamma-ray pulsars}

In addition to searching for gamma-ray pulsations from known radio (or X-ray) pulsars, one of the goals of \emph{Fermi} is to discover new 
pulsars altogether. A "pulsar search consortium" (similar in nature to the "pulsar timing consortium" described above) has been organized for the 
purpose of carrying out follow-up radio observations of any newly-discovered $Fermi$ gamma-ray pulsars. As described in the introduction, one 
of the main objectives of the mission is to carry out population studies of radio-loud and radio-quiet pulsars, and the first step in this 
process is to first discover new radio-quiet (or radio-faint) gamma-ray pulsars.

The term "blind" in these searches refers to the fact that the spin parameters of the potential 
pulsar (e.g. frequency, $f$, and frequency derivative, $\dot{f}$) are unknown\footnote{Fortunately, unlike radio searches, gamma-ray searches need not 
be concerned with searching over a range of values of dispersion measure (DM).}. Although the location is also technically unknown, 
it is taken to be that of an astrophysical source, rather than a random point in the sky. In this respect, a more accurate (though awkward) term 
might be "blind frequency search". Only a relatively small number of locations in the sky is searched (a few hundred in the first few months). The 
arrival times of photons must be corrected to the Solar System barycenter assuming this location, and for each source, a large parameter space 
in $f$ and $\dot{f}$ must be covered. Using the Crab to define our limiting parameters, a maximum frequency of 64 Hz was chosen, and a minimum ratio 
of $\dot{f}$/$f$=-1.25$\times10^{-11}$ s$^{-1}$ (the parameter space is explored in steps of $\dot{f}$/$f$). These parameters cover roughly 95\% of 
the $\sim$2000 pulsars contained in the ATNF database.
Source locations, in general, are derived from the LAT itself. These are mostly new $Fermi$-LAT gamma-ray unidentified sources, though many are coincident with 
old EGRET unidentified sources, only much better localized. However, there is another category of 
source locations used in the pulsar blind searches which correspond to well-studied astrophysical sources which, with various degrees of confidence, are 
suspected of being pulsars. These are often referred to as "Geminga candidates", and a list of such sources was compiled before 
the launch of $Fermi$. In addition to being available straight away (unlike the $Fermi$-LAT locations which take time to generate, and rely on the 
constant accumulation of data to improve such locations), these source locations also have the advantage of being obtained mostly at other 
wavelengths (e.g. X-rays), with much better angular resolution, potentially leading to more sensitive search for pulsations.
At the top of the list of "Geminga candidates" are CTA 1~\cite{2004ApJ...612..398H} and 3EG 1835+59 (aka the 'next' 
Geminga)~\cite{2007ApJ...668.1154H}, both of which benefitted from pointed observations during the comissioning phase (leading to three times as many photons 
accumulated over a similar time period in survey mode). 
In addition to these two, blind searches have been carried out on a number of other sources of interest, such as compact objects in Pulsar 
Wind Nebulae (PWNe), or TeV sources (e.g. those discovered by HESS or by the Milagro experiment~\cite{2007ApJ...664L..91A}). In fact, the "Dragonfly" pulsar 
was independently discovered in a blind search on the location of the TeV source MGRO J2019+37.

\subsubsection{The time-differencing technique}

As described in the previous section, the low fluxes typical of gamma-ray sources result in very long, sparse data sets. The application of 
traditional FFT techniques on such data quickly become unfeasible. 
First, the frequency resolution from such long viewing periods results in FFTs with billions (or tens of billions) of 
frequency bins. More importantly, the relatively large frequency derivatives of pulsars means that these FFTs would have to be computed a large number 
of times in order to span a realistic parameter space, given the small step size in $\dot{f}$ required to keep the signal power within a single FFT bin. 
To make matters worse, the long viewing periods increase the chances that the pulsar will experience a glitch during this time. 
A new technique, known as "time-differencing"~\cite{2006ApJ...652L..49A,2008ApJ...680..620Z}, has been developed, which greatly improves the 
prospects of carrying out these searches. Based on the premise that a periodic signal present in a time series should also show up when analyzing the 
differences of these times, the method involves taking time differences only up to a short time window (on the order of days, rather than months or 
years). In doing this, the required number of FFT bins, $N$, is greatly reduced ($N = 2 \times f \times T_{w}$, where $f$ represents 
the maximum frequency searched and $T_{w}$ is the maximum time-difference window). The reduced frequency resolution also results in a larger step 
size required for $\dot{f}$, greatly reducing the number of $\dot{f}$ trials needed. The overall result is that the computational 
and memory costs become a fraction of the standard FFT methods, with only a small reduction in sensitivity. 
This makes it possible to carry out a large number of blind searches over a realistic parameter space on standard desktop computers with only a 
few GB of memory. That said, the blind search efforts have been significantly enhanced by access to the UCSC Astrophysics Supercomputer Pleiades, 
with 207 Dell PowerEdge 1950 compute nodes (828 processing cores). Despite the efficiency of the time-differencing technique, it is still 
desirable to limit the number of source locations in the sky that are searched, given both the finite computational resources available, and also to avoid 
incurring large penalties in the significance of our detections, due to the increasing number of trials used in finding a particular signal.

\section{A radio-quiet pulsar in CTA 1}

\begin{figure}
  \includegraphics[height=.3\textheight]{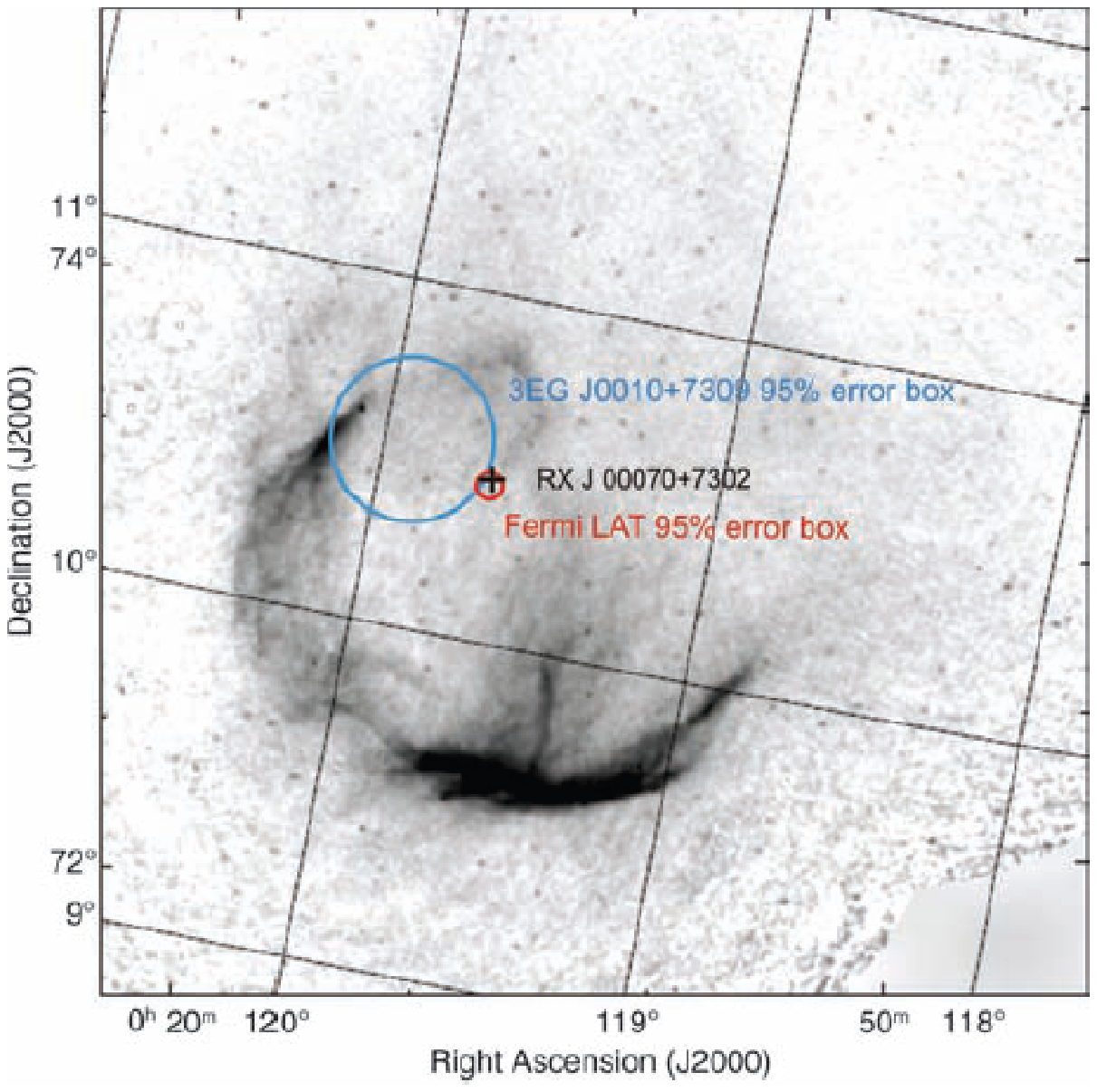}
  \includegraphics[height=.3\textheight]{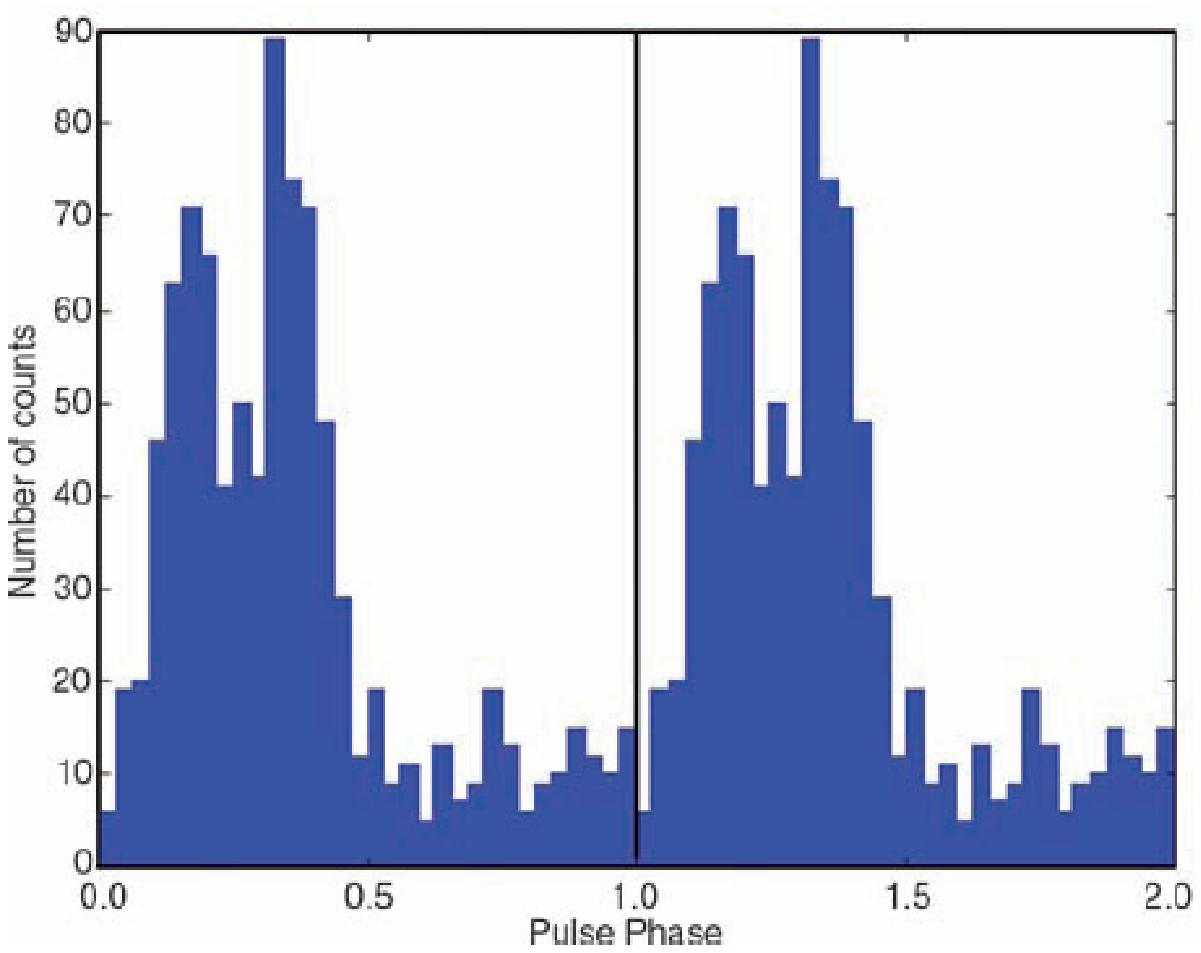}
\caption{
{\bf Left} -- \emph{Fermi}-LAT gamma-ray source location of CTA 1, superimposed on a 1420-MHz radio image of the CTA 1 SNR. The red circle shows the 
\emph{Fermi}-LAT 95\% containment radius, while the cross represents the location of the X-ray point source. The large blue circle shows the corresponding
EGRET 95\% error circle (taken from ~\cite{2008Sci...322.1218A}).
{\bf Right} -- Gamma-ray ($>$ 100 MeV) folded light curve of the CTA 1 pulsar shown with two periods of rotation and 32 time bins per period
(from ~\cite{2008Sci...322.1218A}).}
\label{fig7}
\end{figure}

The first major $Fermi$ discovery (now published in $Science$) came with the detection, in a blind search, of pulsations from CTA 1. This young, nearby, shell-type SNR was 
discovered in radio in the 1960s and X-ray observations show a well-localized central point source, RXJ0007.0+7303, embedded in a pulsar wind nebula 
(PWN)~\cite{2004ApJ...612..398H}. High energy ($>$ 100 MeV) emission was detected by EGRET from 3EG J0010+7309, coincident with this source 
(see Figure~\ref{fig7}), but the low number of photons and poor localization made the search for pulsations very challenging. 
The new pulsar, with a period of 315.8637050 ms and a period derivative of 3.615$\times$10$^{-13}$ s s$^{-1}$ is a typical young, energetic
pulsar, with a derived characteristic age of $\sim$14,000 years (consistent with the estimated age of the SNR) and a spin-down power of 
4.5$\times$10$^{35}$ erg s$^{-1}$. For more details, see \cite{2008Sci...322.1218A}.

\section{Conclusions}
The $Fermi$ Gamma-ray Space Telescope has been performing flawlessly since its launch last June. One of its key scientific missions is to 
resolve the gamma-ray sky, identifying many (if not all) the gamma-ray sources that its predecessor, EGRET, left 
unidentified. Many of these EGRET unidentified sources are expected to be pulsars. The hugely improved sensitivity and angular resolution of \emph{Fermi}, 
combined with a new powerful time-differencing technique, makes it, for the first time, possible to carry out sensitive searches for gamma-ray pulsations 
from all these sources. 
Even before the commissioning phase was complete, \emph{Fermi} has already yielded important scientific discoveries, including the detection 
of gamma-ray pulsations from the "Dragonfly" pulsar (PSR J2021+3651), the detection of the first gamma-ray glitch from PSR B1706--44, and the discovery of a 
new radio-quiet gamma-ray pulsar, in the SNR CTA 1. This last discovery, published in $Science$, represents a major milestone, as it is the first time that 
a pulsar has been discovered using gamma-ray observations alone. Based on these early results, \emph{Fermi} is already fulfilling its pre-launch expectations 
and is beginning to pave the way for a new era in gamma-ray astrophysics in general, and pulsar astrophysics in particular.
.


\begin{theacknowledgments}

I want to thank the organizers for putting together a wonderful conference, where 
the quality of the talks was matched only by that of the food. In particular, I am grateful 
to Denis Bastieri and Riccardo Rando for all their work and hospitality during the conference, as well as for 
their efforts (and patience) in the preparation of this volume.  

The $Fermi$ LAT Collaboration acknowledges the generous support of a number of 
agencies and institutes that have supported the $Fermi$ LAT Collaboration. These 
include the National Aeronautics and Space Administration and the Department of 
Energy in the United States, the Commissariat \`a l'Energie Atomique and the 
Centre National de la Recherche Scientifique / Institut National de Physique 
Nucl\'eaire et de Physique des Particules in France, the Agenzia Spaziale Italiana, the Istituto Nazionale 
di Fisica Nucleare (INFN), and the Instituto Nazionale di Astrofisica (INAF) in Italy, the Ministry of 
Education, Culture, Sports, Science and Technology (MEXT), High Energy 
Accelerator Research Organization (KEK) and Japan Aerospace Exploration Agency 
(JAXA) in Japan, and the K. A. Wallenberg Foundation, the Swedish Research Council, and the Swedish National 
Space Board in Sweden. I am grateful for the support of the American Astronomical Society and the National 
Science Foundation in the form of an International Travel Grant, which enabled me to attend 
this conference.

Finally, I would like to dedicate this article to the memory of my brother, Carlos Roberto Saz Parkinson.


\end{theacknowledgments}



\bibliographystyle{aipproc}   


\IfFileExists{\jobname.bbl}{}
 {\typeout{}
  \typeout{******************************************}
  \typeout{** Please run "bibtex \jobname" to optain}
  \typeout{** the bibliography and then re-run LaTeX}
  \typeout{** twice to fix the references!}
  \typeout{******************************************}
  \typeout{}
 }

\def \apjl {ApJL}
\def \apj {ApJ}

\bibliography{ms}

\end{document}